# Novel Potassium Polynitrides at High Pressures

Brad A. Steele and Ivan I. Oleynik*

*Department of Physics, University of South Florida, 4202 East Fowler Ave., Tampa, FL 33612*

E-mail: oleynik@usf.edu

**Abstract**

Polynitrogen compounds have attracted great interest due to their potential applications as high energy density materials. Most recently, a rich variety of alkali polynitrogens ($R_xN_y$; R=Li, Na, and Cs) have been predicted to be stable at high pressures and one of them, $CsN_5$ has been recently synthesized. In this work, various potassium polynitrides are investigated using first-principles crystal structure search methods. Several novel molecular crystals consisting of $N_4$ chains, $N_5$ rings, and $N_6$ rings stable at high pressures are discovered. In addition, an unusual nitrogen-rich metallic crystal with stoichiometry $K_2N_{16}$ consisting of a planar two-dimensional extended network of nitrogen atoms arranged in fused eighteen atom rings is found to be stable above 70 GPa. An appreciable electron transfer from K to N atoms is responsible for the appearance of unexpected chemical bonding in these crystals. The thermodynamic stability and high pressure phase diagram is constructed. The electronic and vibrational properties of the layered polynitrogen $K_2N_{16}$ compound are investigated, and the pressure-dependent IR-spectrum is obtained to assist in experimental discovery of this new high-nitrogen content material.

# Introduction

At ambient conditions nitrogen forms one of the strongest triple bonds found in nature in the diatomic molecule ($N_2$). Upon compression to high pressures ($>$ 100 GPa) and temperatures ($>$ 1000 K) the $N_2$ molecular crystal undergoes the chemical transformation into the extended single-bonded cubic gauche phase of nitrogen[1] or the layered polymeric phase of nitrogen[2]. These single-bonded extended crystals are high energy density materials with an energy density much larger than conventional explosives[3]. The energy density comes from the large energy difference between the strengths of the nitrogen single and triple bonds, as well as a larger density of dense polymeric nitrogen in comparison with sparse molecular crystals. However, recovery of cubic gauche or the layered polymeric phases of nitrogen at ambient conditions has proven to be challenging[1,2,4,5].

By incorporating metallic impurities into the crystal structure, a charge transfer from metallic to nitrogen atoms occurs, which can enhance the stability of nitrogen oligomers, while introducing an ionic bonding between metal cations and the nitrogen oligomer anions[6–12]. Recent investigations of alkali polynitrides ($R_xN_y$; R=Li, Na, and Cs) have revealed a rich variety of stable all-nitrogen oligomers at high pressures including $N_4$ chains, $N_5$ rings, $N_6$ rings, and infinite nitrogen chains[7–12]. In fact, the elusive $N_5^-$ anion with the bond strength between single and double N-N bond, was recently synthesized as a salt with cesium cations by compressing cesium azide ($CsN_3$) in a nitrogen-rich environment to 60 GPa[12]. Non-metallic compounds, ammonium pentazolate $NH_4N_5$ and long-sought pentazole $HN_5$, containing $N_5^-$ have also been recently discovered[13]. In addition, alkali single-bonded extended structures of nitrogen have been predicted to be stable in stoichiometries $KN_3$ above 299 GPa[7], $LiN_2$ above 60 GPa[10], $CsN_2$ above 40 GPa[9], and $NaN_2$ above 50 GPa[11].

How the nature of a specific group-I alkali metal atom affects the chemical bonding within the family of alkali metal polynitrides is an interesting question worth investigating. Different electronegativities and ionic radii are expected to affect the amount of charge transfer between alkali cations and nitrogen anions. Therefore, it is quite plausible that in addition to observing common crystalline compounds, such as alkali pentazolates and those containing infinite linear chains and $N_6$ hexazine rings, new crystals with unusual bonding patterns specific to potassium polynitrides do exist.

In this work, potassium polynitrides ($K_xN_y$) are systematically investigated by performing variable composition crystal structure search of compounds composed of potassium and nitrogen at a range of pressures from 0 to 100 GPa. The goal is to discover novel single-bonded polymeric forms of nitrogen that might be different from known polynitrogen compounds in other alkali polynitrides investigated previously. These new compounds are theoretically characterized by calculating electronic and vibrational properties using first-principles density functional theory.

# Methods

The search for new nitrogen-rich potassium polynitrogen compounds of varying stoichiometry is performed at 30, 60 GPa and 80 GPa using first-principles evolutionary crystal structure prediction method USPEX[14–16]. The USPEX methodology of evolutionary crystal structure prediction works as follows. During variable composition calculations at a given pressure, USPEX creates a specified number of structures at each generation step, which contain both newly randomly generated structures as well as those produced from the lowest formation enthalpy structures from a previous generation. In the first generation, every structure is randomly generated by random sampling of the stoichiometry, crystal symmetry, ionic coordinates, and lattice parameters. At subsequent generations, physically motivated variation operators are applied to the best structures from the previous generation to produce new structures for the new generation. Each structure is optimized at a given pressure using first-principles code VASP[17] by varying the ionic coordinates and lattice parameters to achieve minimum enthalpy. The optimized structures are then ranked by enthalpy and the lowest enthalpy structures "survive", i.e. passed to the next generation. The search is stopped after the lowest energy structures do not change after 5-10 generations. More information about structure prediction can be found in Refs.[14,16,18].

The initial variable composition search is performed using 8-20 atoms/unit cell and its quality is verified by predicting the known phases of potassium azide[7,19,20]. After the variable composition search is completed, fixed-compositions structure searches with larger number of atoms (up to 24 atoms/unit cell) are performed to find the lowest enthalpy structure for each composition.

During the structure search, the cell parameters as well as the atomic coordinates of each structure are optimized at a given pressure to minimize its enthalpy using the Perdew-Burke-Ernzerhof (PBE) generalized gradient approximation (GGA)[21] within density functional theory (DFT) as implemented in VASP[17]. The dispersive correction due to Grimme[22] is added to the DFT energy, forces, and stresses to take into account the long-range van der Waals interactions. A comparison of the calculated lattice parameters to experiment for potassium azide crystal at ambient conditions given in Table 1 demonstrates that PBE+vdW functional with vdW correction gives lattice parameters in good agreement with experiment (error being less than 1.1 %), thus providing justification for using PBE+vdW in DFT calculations. For the structure search, projector augmented wave (PAW) pseudopotentials[23] and plane wave basis sets are used with an energy cutoff of 430 eV and a 0.07 $Å^{-1}$ k-point sampling. After the structure search is completed, more accurate calculations are performed to compute the convex hulls using hard PAW pseudopotentials for N with inner core radii of 0.582 Å for N, and energy cutoff of 1,000 eV and 0.05 $Å^{-1}$ k-point sampling. The vibrational properties are calculated within the frozen phonon approximation. The IR frequencies and absolute intensities are obtained by computing phonon modes at the Γ-point together with the corresponding derivatives of the macroscopic dielectric tensor with respect to the normal mode coordinates[24,25]. Charges on atoms and bond orders are calculated using LCAO code DMol[26]

Table 1: Theoretical lattice parameters calculated using PBE GGA and PBE plus vdW D2 corrections (PBE+vdW) compared with experiment for potassium azide $KN_3$-I4/mcm at ambient conditions.

| Parameter | Experiment[19] | PBE+vdW | PBE |
|---|---|---|---|
| a=b(Å) | 6.1109 | 6.1177 (0.1 %) | 6.2297 (1.9 %) |
| c(Å) | 7.0975 | 7.0051 (-1.3 %) | 7.2093 (1.6 %) |
| V(Å$^3$) | 265.043 | 262.11 (-1.1 %) | 279.75 (5.5 %) |

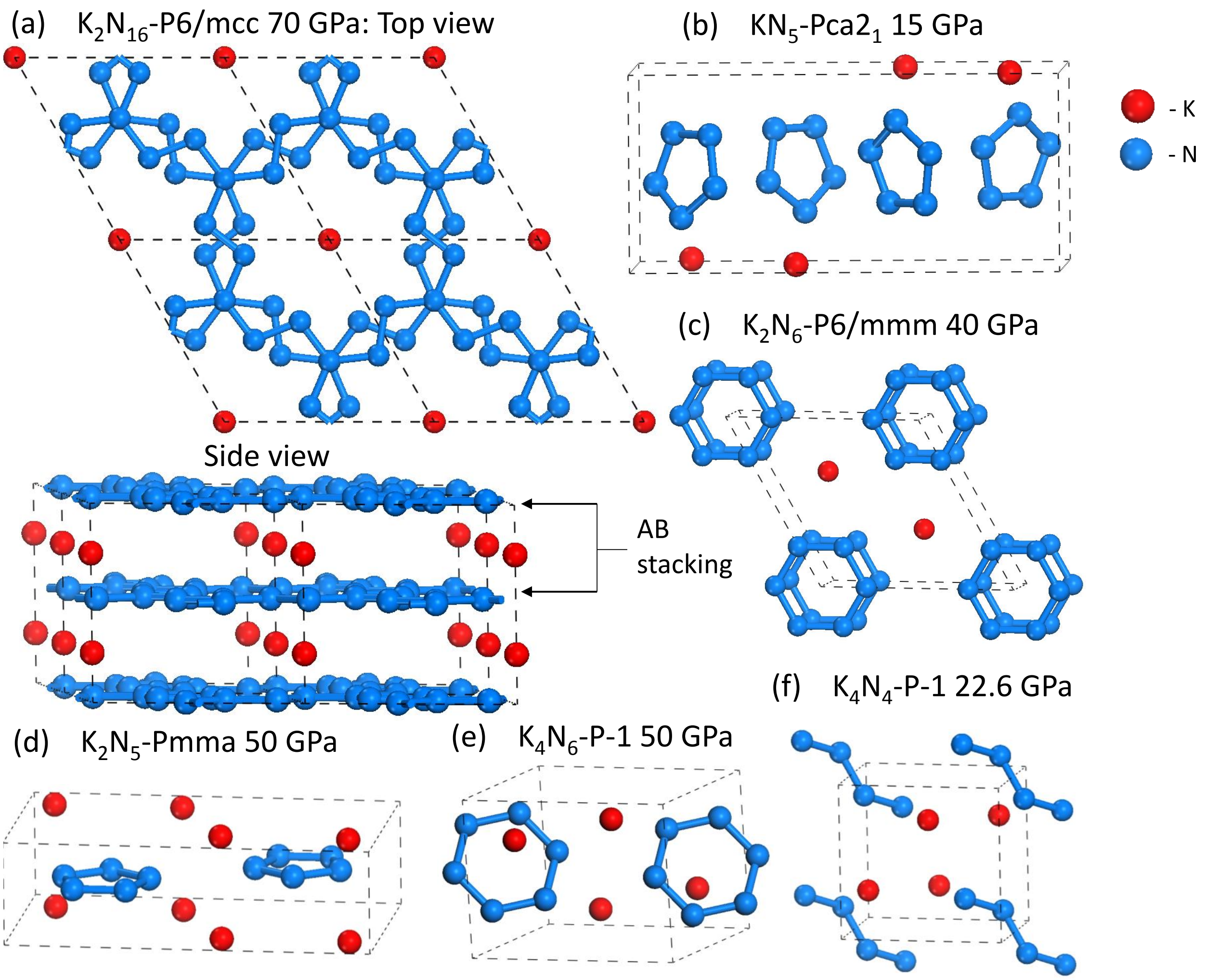

Figure 1: Snapshots of several potassium polynitrides discovered in the variable composition $K_xN_y$ structure search at the pressures they become stable. (a) The novel extended hexagonal $K_2N_{16}$ compound, containing layered polynitrogen, stable above 70 GPa shown in the ab plane and ac plane. (b) Crystal structure of $KN_5$ stable above 15 GPa, (c) Crystal structure of $K_2N_6$ at 40 GPa, (d) Crystal structure of $K_2N_5$ at 50 GPa, (e) Crystal structure of $K_4N_6$ at 50 GPa, and (f) Crystal structure of $K_4N_4$ at 20 GPa.

# Results and discussion

The structure search results in the discovery of several polynitrogen species stable at high pressures, see Figure 1(a-f). The following compounds containing the nitrogen components are found (listed from the most N-rich to the least N-rich stoichiometries and labeled by stoichiometry followed by space group symmetry): (1) $K_2N_{16}$-P6/mcc crystal, consisting of planes of fused 18 atom rings forming an extended planar network of N atoms and K atoms in between the nitrogen planes (Figure 1 (a)); (2) $KN_5$-Pca$2_1$ crystal, which consists of $N_5$ pentazole rings (Figure 1(b)), (3) $K_2N_6$-P6/mmm crystal, which consists of $N_6$ hexazine rings (Figure 1 (c)), (4) $K_2N_5$-Pmma crystal, which consists of $N_5$ pentazole rings (Figure 1 (d)); (5) $K_4N_6$-P-1 crystal, which consists of $N_6$ hexazine rings (Figure 1 (e)) similar to what is found in case of Cs[12]; (6) $K_4N_4$-P-1 crystal consisting of $N_4$ chains (Figure 1 (f)) also similar to what is found in case of Cs[9,12]. The calculated convex hulls at pressures 15, 30, 60 and 80 GPa are shown in Figure 2(a), whereas the phase diagram for all predicted structures is shown in Figure 2(b).

Interestingly, the compounds with 1:1 stoichiometry undergo some irregular phase transformations as pressure increases: the $K_2N_2$-Cmmm crystal, containing $N_2^-$ anions, is on the convex hull and is stable from 10 GPa up to 22.6 GPa. From 22.6 GPa up to 41.7 GPa the $K_4N_4$-P-1 crystal containing $N_4^-$ anions (Figure 1(f)) is the lowest enthalpy structure for the 1:1 stoichiometry, see Figure 2(b) for the phase diagram and the Supporting Information Figure S1 for the enthalpy difference between these two phases. At pressures higher than 41.7 GPa the $K_2N_2$-Cmmm molecular crystal, containing $N_2$ anions once again is the lowest enthalpy structure with the 1:1 ratio of K:N.

Since all alkali metals are monovalent, it is expected that the alkali polynitrides will be very similar. It turns out not to be the case: we find that there are several important differences between alkali polynitrides that have been predicted thus far. A summary of the alkali polynitrogen compounds that have been theoretically predicted are given in Table 2. Potassium polynitride with the stoichiometry $R_2N_{16}$ is the only alkali metal containing compound which consists of planes of nitrogen atoms arranged in a planar network of fused 18-atom rings and K atoms in between the N planes, see Figure 1(a) and Table 2. In contrast, alkali pentazolates $RN_5$ exist as stable compounds at high pressures for all alkali (R) metals. Alkali azides are typically predicted to transform into a crystal structure containing $N_6$ anions with either the P6/mmm space group for potassium[7,27] or the P6/m space group for lithium[8,10]. Our calculations show that $K_2N_6$-P6/mmm crystal is on the convex hull, i.e. thermodynamically stable, above 40 GPa. A similar crystal structure was predicted for Na[28] as well, but our previous results indicate it is not on the convex hull and is therefore metastable[11], see Table 2. In contrast to the case of $CsN_2$[9], the crystal structure of $KN_2$-C2/c compound consisting of nitrogen chains is calculated to be unstable, see Figure 2(a). A different crystal structure with the same stoichiometry is predicted to be stable for Na[11]. It is expected that potassium and heavier alkali metals may exhibit more complex chemistry due to an increased contribution from semi-core d-oribitals, which also leads to complex post-fcc alkali metallic structures[29]. As can be seen in Table 2 every stoichiometry encountered for other alkali polynitrides occur also for potassium and cesium polynitrides, with the exception of just one stoichiometry, $RN_2$. The $R_4N_4$ and the $R_4N_6$ stoichiometries are not stable for either sodium (R=Na) or lithium (R=Li), but $R_2N_6$ which is similar to $R_4N_6$ is stable for lithium (R=Li) but not for sodium (R=Na). These results demonstrate that the nature of the alkali metal constituents affects in a significant way the high pressure chemistry of alkali polynitrides.

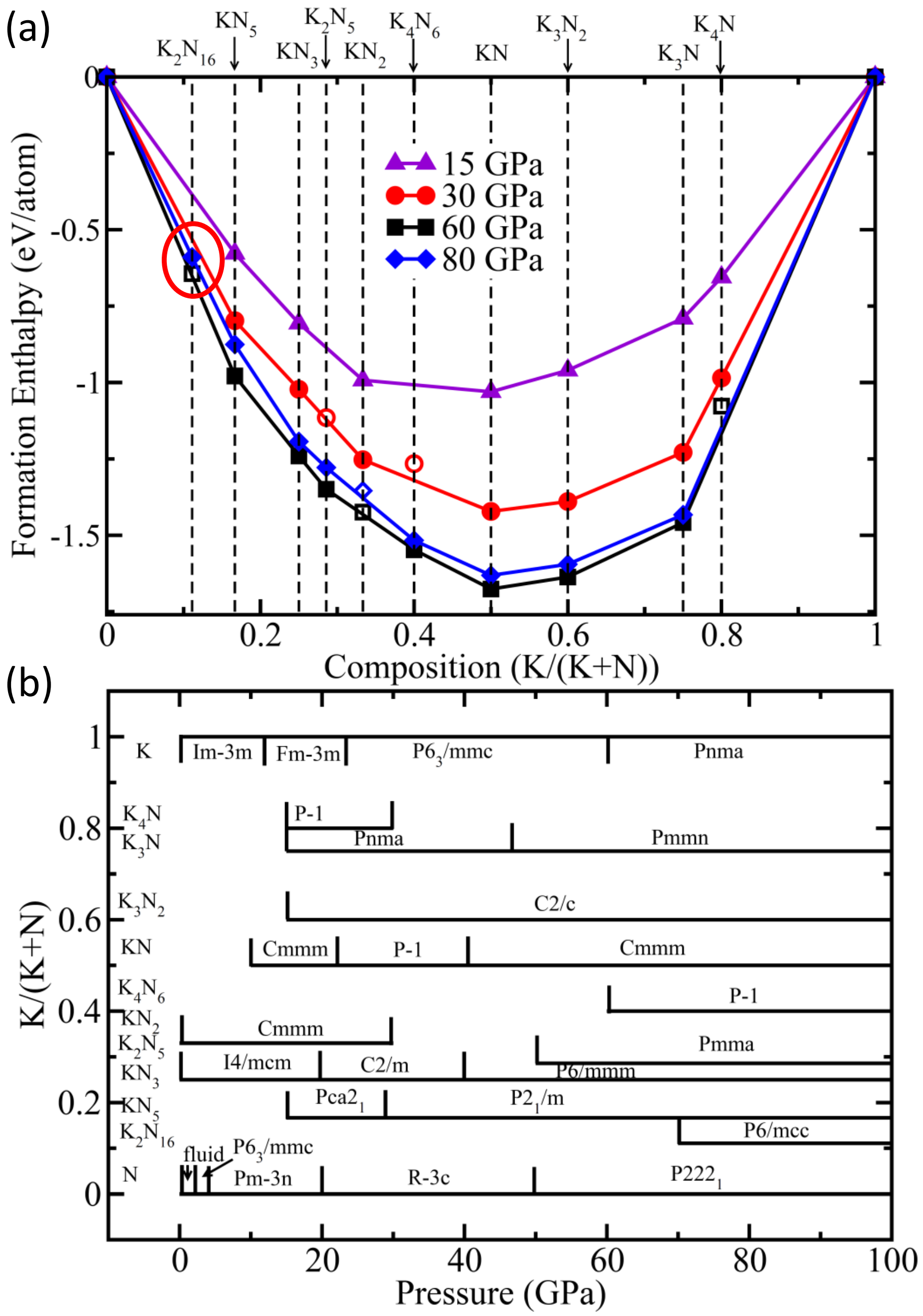

Figure 2: (a) $K_xN_y$ convex hull diagram from 15-100 GPa. Solid symbols are stable and open symbols – metastable phases. (b) Theoretical phase diagram for $K_xN_y$.

Table 2: Comparison of the stoichiometries and corresponding nitrogen oligomers for thermodynamically stable nitrogen-rich alkali polynitrides in the interval of pressures from 0 at to 100 GPa. The data for $K_xN_y$ are from this work, $Li_xN_y$ - from Ref.[10], $Na_xN_y$ - from Ref.[11], and $Cs_xN_y$- from Refs.[9,12].

| Stoichiometry ($R_xN_y$) | Alkali Metal | N oligomer | Pressure |
|---|---|---|---|
| $R_2N_{16}$ | K | fused $N_{18}$ rings | 70 GPa |
| $RN_5$ | Li, Na, K, Cs | $N_5^-$ pentazole rings | ~15 GPa |
| $R_2N_6$ | Li, K, Cs | $N_6^-$ hexazine rings | 40-80 GPa |
| $R_2N_5$ | Na, K, Cs | $N_5^-$ pentazole rings | 30-50 GPa |
| $RN_2$ | Li, Na, Cs | N chains | 50-60 GPa |
| $R_4N_6$ | K, Cs | $N_6^-$ hexazine rings | 30-60 GPa |
| $R_4N_4$ | K, Cs | $N_4^-$ chains | 20-45 GPa |

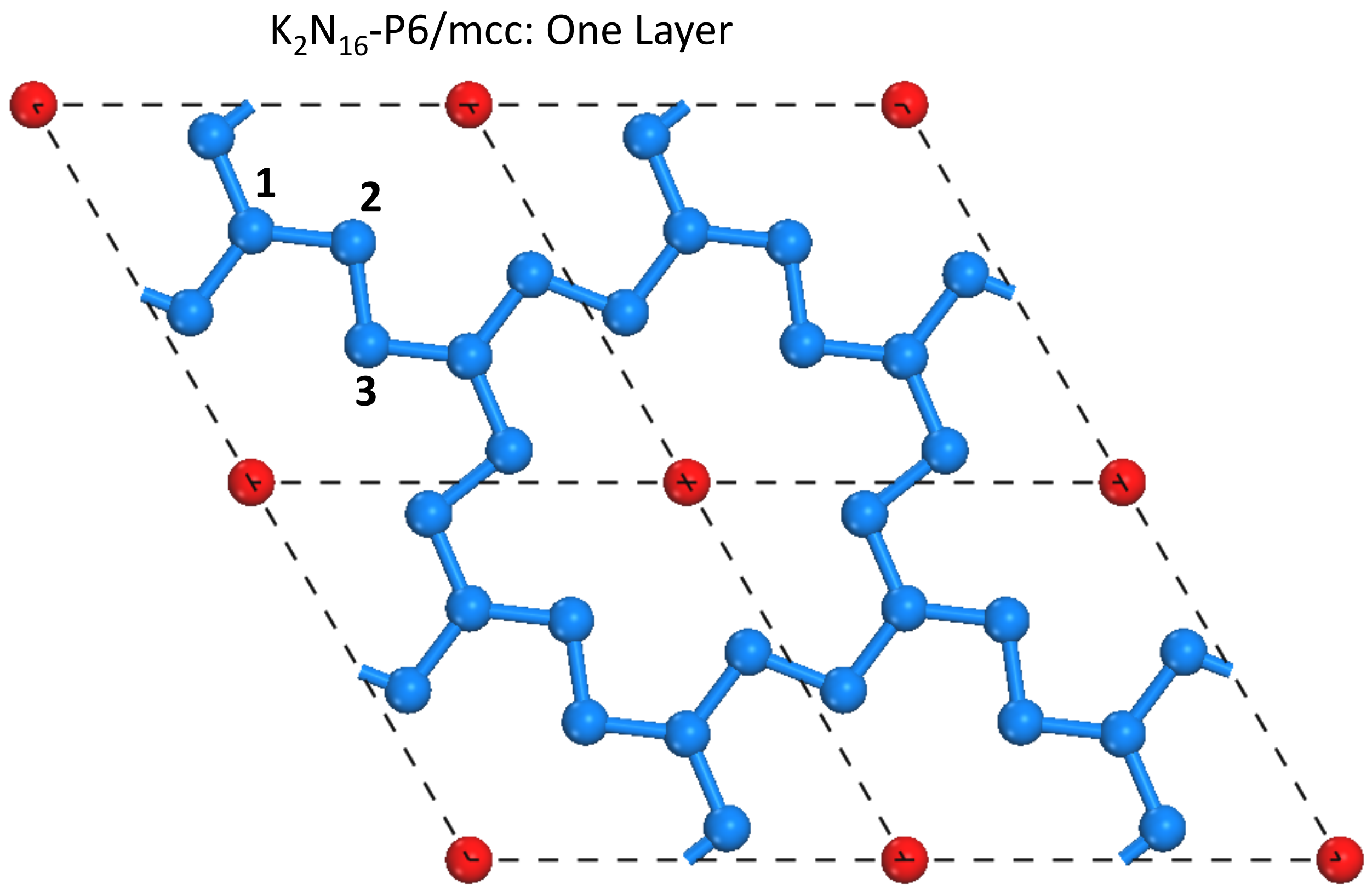

Figure 3: Crystal structure of one layer of $K_2N_{16}$-P6/mcc compound, consisting of a planar network of fused $N_{18}$ rings (N atoms are blue colored) plus K atoms in the corners of the unit cell above and below N-planes.

The most interesting structure discovered during the search is the extended layered nitrogen-containing crystal that consists of nitrogen planes with N atoms forming planar infinite network of fused $N_{18}$ rings, see Fig. 3. This high-nitrogen content crystal with stoichiometry $K_2N_{16}$ and hexagonal space group symmetry P6/mcc is predicted to be stable above 70 GPa, see Figure 1(a). Although 70 GPa is even higher than the predicted stability pressure of cg-N of 50 GPa[30] it is still of fundamental significance to uncover novel forms of extended solid nitrogen in high-nitrogen content compounds as such unusual nitrogen structures might exist at lower and even ambient pressures in compounds containing elements other than potassium. Each $N_{18}$ ring spans four adjacent unit cells, each containing eight N atoms: two of them are three-fold coordinated at the junction point of two neighboring $N_{18}$ rings, the other six N atoms are two-fold coordinated, see Figure 3. The entire crystal structure is layered and consists of atomic planes: the first plane is the layer of interconnected $N_{18}$ rings, the second contains potassium cations in the corners of the unit cell, the third plane is the layer of interconnected $N_{18}$ rings again but stacked differently in respect to the first N plane. The layered structure of $K_2N_{16}$-P6/mcc crystal can be represented as having A$\beta$ C$\beta$ stacking, where A and C label the nitrogen planes, and $\beta$ – K planes. This structure is on the convex hull at 70 GPa up to 100 GPa (this highest pressure studied), see Figure 2(a). The dynamical stability is evaluated by calculating the phonon dispersion curves shown in Figure 4 at 80 GPa with no imaginary frequencies in the entire Brillioun zone. At ambient pressure the AB stacking of nitrogen planes is calculated to be dynamically unstable, however the AA stacking of nitrogen planes is 22 meV/atom lower in energy and contains no imaginary frequencies. Therefore the AA stacking of nitrogen planes is dynamically stable, see Supporting Information Figure S2. The structure is also metallic with occupied bands at the Fermi level, see Figure 5(b).

The electronic band structure of the $K_2N_{16}$-P6/mcc crystal features relatively flat bands near the Fermi level especially those along the Brillouin zone (BZ) direction corresponding to real space direction parallel to the layers (planes) of atoms, e.g. along A to H high symmetry direction of the BZ, see Figure 5(b). This causes a small peak at the Fermi level, see Figure 5(b). These bands are almost exclusively nitrogen p-orbitals which participate in the N-N bonding in the plane of nitrogen atoms. The nitrogen planes consist mostly of N-N single bonds, which have a bond order of 1.15 for every N-N bond other than the bond between atoms N(2) and N(3) shown in Figure 3, that have a bond order of 1.36, see Table 3. The three-fold coordinated single bonded atoms form an $sp^2$ geometry in the plane, see Figure

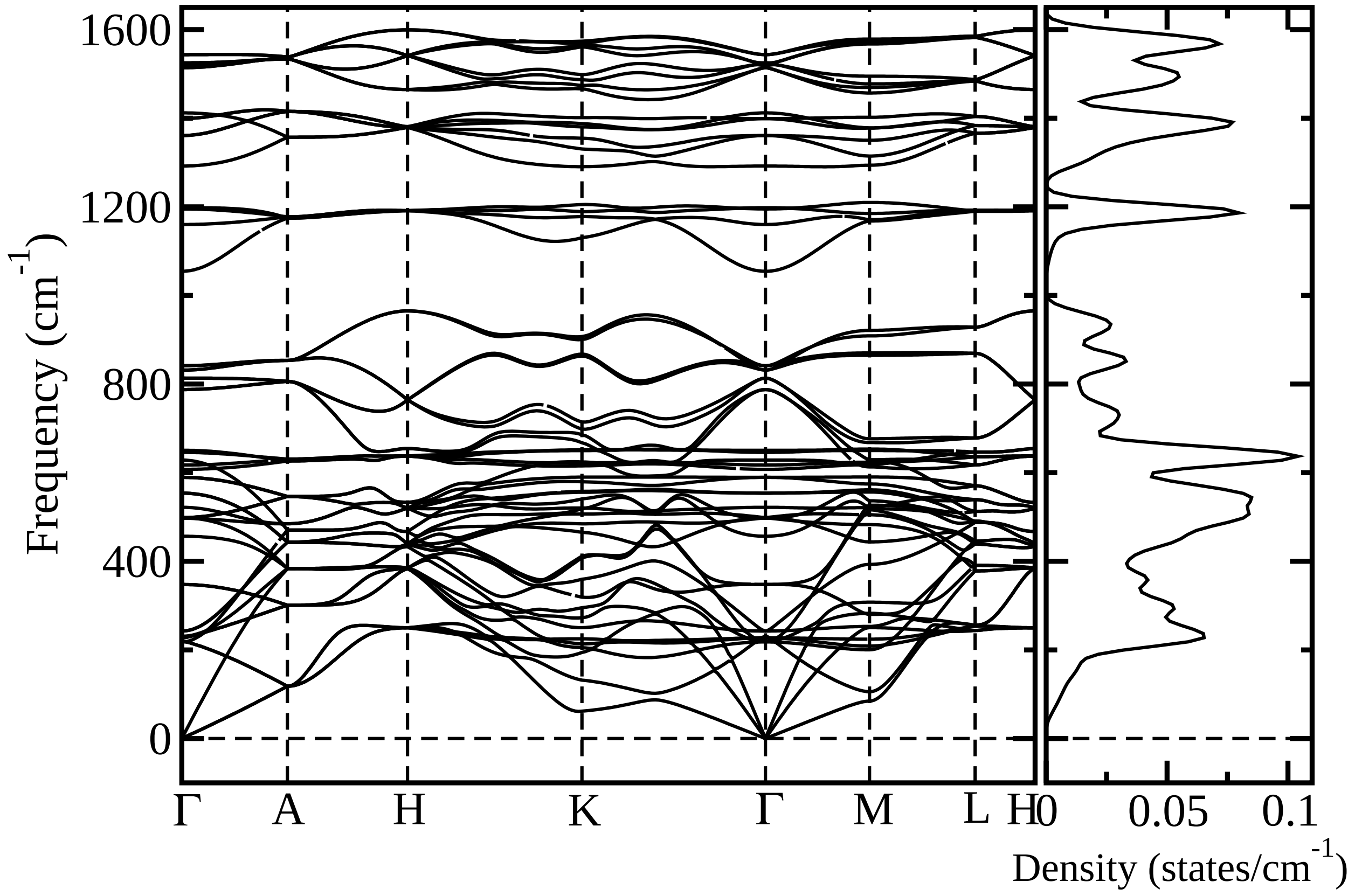

Figure 4: Phonon dispersion curves for extended layered $K_2N_{16}$ at 80 GPa showing no imaginary frequencies in the entire Brillioun zone.

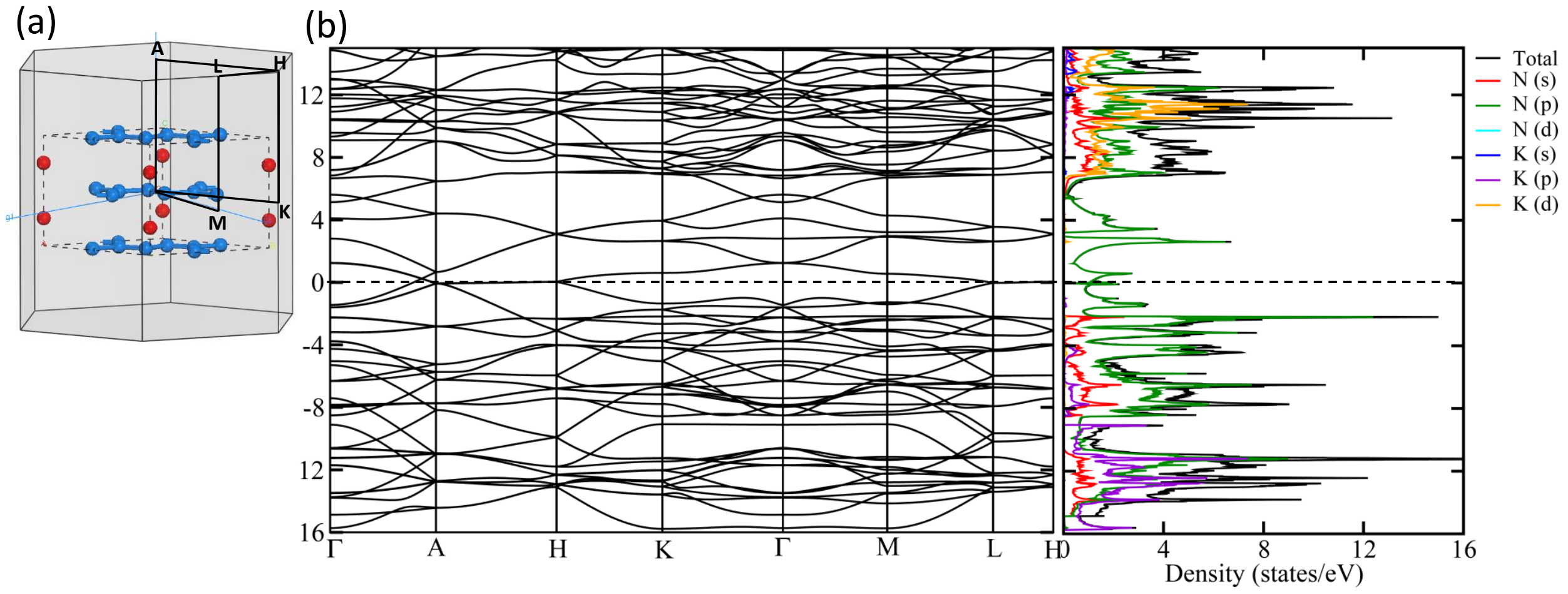

Figure 5: (a) The Brillouin zone (BZ) of the hexagonal unit cell of $K_2N_{16}$-P6/mcc with high-symmetry k points labeled to plot the band structure along high symmetry directions of the BZ. (b) electronic band structure of $K_2N_{16}$-P6/mcc crystal at 80 GPa. The total and local density of states are also plotted.

3. It can be seen from the local density of states given in Figure 5(b), that there are several states occupied by both s and p N orbitals thus confirming the $sp^2$ nature of the N-N single bonds. Atoms N(2) and N(3) are two-fold coordinated which give them a slightly higher bond order. Interestingly atom N(1) in the N-network carries a positive charge of 0.168 whereas the charges on the rest of the nitrogen atoms are $-0.178$. There is certainly a net transfer of negative charge to N atoms from K atoms, the charge on each potassium atom being +0.731. The charges and bond orders are summarized in Table 3. Interestingly, K p orbitals, and even d orbitals at higher energies, contribute appreciably to the density of states, see Figure 5. A contribution from d-orbitals also plays a role in producing the unusual phase diagram of pure K at high pressures[29].

The charges and bond orders for the other potassium polynitrides given in the Table 4 are similar to those found in analogous crystals of other alkali polynitrides. The bond orders for most polynitrogen species are slightly larger than 1 except for potassium pentazolate ($KN_5$), which has a bond order of 1.41, see Table 4. This is because of aromatic nature of the bond in the pentazole that increases the strength of the bond. The $K_2N_6$ and $K_4N_6$ stoichiometries have conjugated N-N bonds with a bond order close to 1 and a total charge less than 1 on the $N_6$ ring, see Table 4. The $K_4N_6$ stoichiometry has the largest charge transfer from potassium to nitrogen atoms with a total charge less than 2 on the $N_6$ ring. The large amount of ionic bonding makes this structure stable even though the bond order is the lowest of the polynitrogen species predicted in this work, see Table 4.

Table 3: The charges and bond orders for nitrogen atoms/bonds in the $K_2N_{16}$-P6/mcc structure where the atom numbers refer to those in Figure 3.

| Atom | Charge | Bond Order (bond with previous atom) |
|---|---|---|
| 1 | 0.168 | NA |
| 2 | -0.178 | 1.15 |
| 3 | -0.178 | 1.36 |

Table 4: Average Mulliken charges and Mayer bond orders for potassium polynitrogen compounds at 0 GPa discovered during the structure search at high pressures.

| Structure | Nitrogen molecule | Charge on N atom | Bond Order |
|---|---|---|---|
| $KN_5$-Pca$2_1$ | $N_5$ ring | -0.17 | 1.41 |
| $K_2N_6$-P6/mmm | $N_6$ ring | -0.23 | 1.19 |
| $K_2N_5$-Pmma | $N_5$ ring | -0.27 | 1.16 |
| $K_4N_6$-P-1 | $N_6$ ring | -0.41 | 1.05 |
| $K_4N_4$-P-1 | $N_4$ chain | -0.37 | 1.18 |

The extended layered material $K_2N_{16}$-P6/mcc can potentially be synthesized by compressing potassium azide with $N_2$ above 70 GPa in the following reaction: $2KN_3+5N_2 \rightarrow K_2N_{16}$. The relative enthalpies for the proposed reaction and for other possible reactions involving potassium polynitrides as a function of pressure are given in Figure 6. First, the mixture of $2KN_3+5N_2$ is predicted to transform into $2KN_5+3N_2$ at 10.4 GPa, then to $K_2N_{16}$-P6/mcc at 59.6 GPa. Kinetic barriers will significantly delay the phase transformations so that larger pressures would likely be required for synthesis. At 60 GPa, all three compounds $K_2N_{16}$-P6/mcc, $K_2N_6$-P6/mmm, and $KN_5$-P$2_1$m are lower in enthalpy than the precursor, but $K_2N_{16}$-P6/mcc is the lowest enthalpy structure at 60 GPa.

The calculated IR-spectra of $K_2N_{16}$-P6/mcc from 57 GPa to 102 GPa are given in Figure 7. The spectrum contains two lattice modes marked with a 'T' in Figure 7. The spectrum also has one out-of-plane and four in-plane N-N stretching modes with the highest frequency at 57 GPa being about 1,410 $cm^{-1}$, see Figure 7. Each mode monotonically blue shifts with pressure.

# Conclusions

Systematic investigation of the high-pressure chemistry of potassium polynitride compounds with variable $K_xN_y$ stoichiometries has been performed using first-principles evolutionary crystal structure search method. Several new crystal structures are discovered with molecular building blocks similar to those that have been predicted for other alkali polynitrides including $N_4$ chains, $N_5$ pentazole rings, and $N_6$ hexazine rings. One new crystal that makes the case of potassium polynitrides interesting is a layered $K_2N_{16}$-P6/mcc crystal consisting of nitrogen planes of fused $N_{18}$ rings sandwiching the potassium planes. This unusual compound is shown to be dynamically stable at 80 GPa and is also predicted to be metallic. The nitrogen planes contain nitrogen bonds with bond orders 1.15 and 1.36 indicating that nitrogen is single bonded. The bond orders and charges for other potassium polynitrides have also been calculated. The IR-spectrum of $K_2N_{16}$-P6/mcc is calculated as a function of pressure and mode assignments were made. The $K_2N_{16}$-P6/mcc crystal is predicted to become more energetically favorable than potassium azide plus di-nitrogen precursors above 41.2 GPa, as well as more favorable than another polynitride, potassium pentazolate $KN_5$, plus di-nitrogen above 59.6 GPa. The discovery of this crystal demonstrates that new and interesting nitrogen-rich crystal structures can exist in extreme environments and should inspire the search for chemically similar compounds containing such unusual extended network of nitrogen atoms.

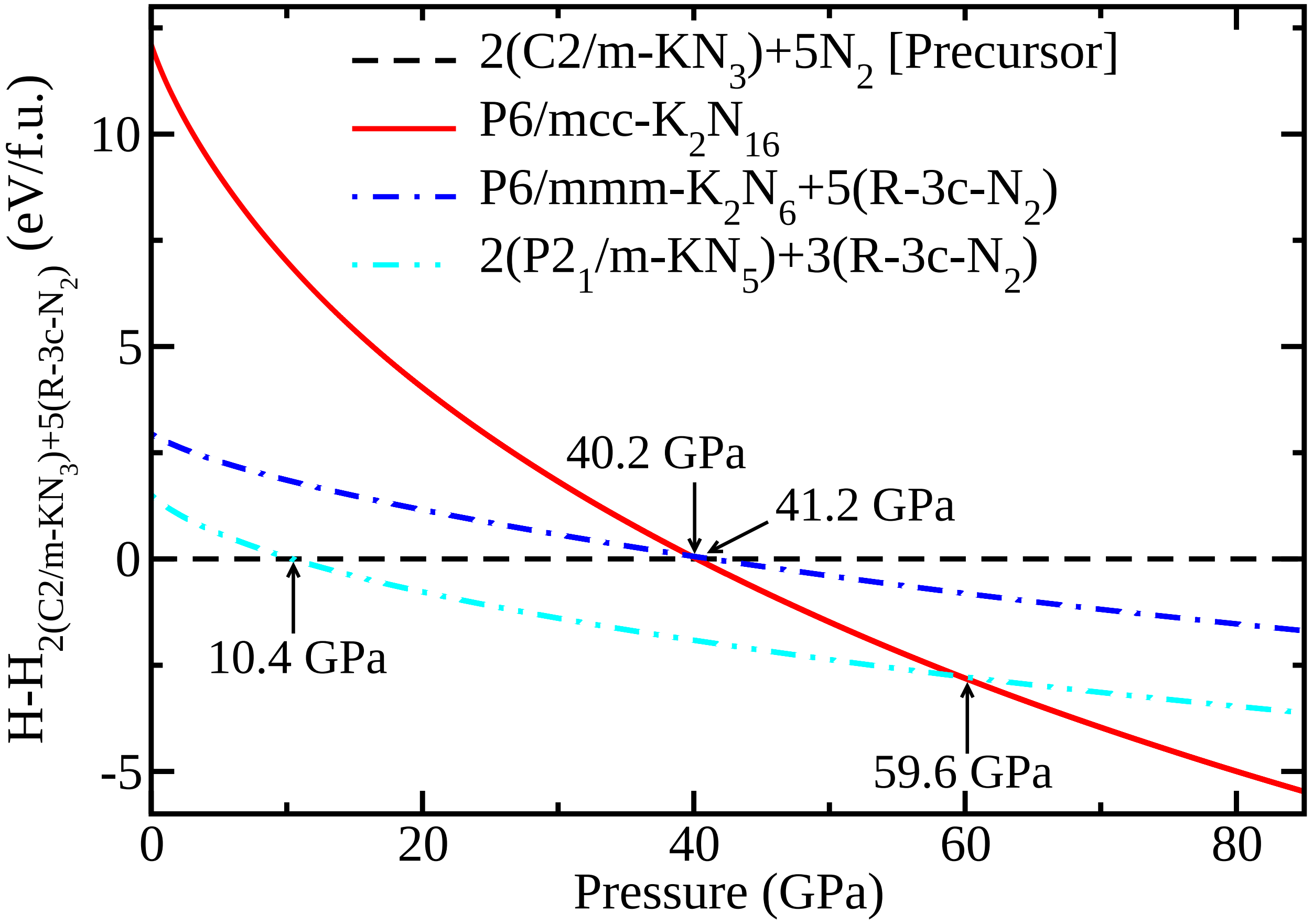

Figure 6: Relative enthalpies and phase transition pathway for the proposed precursor materials $2KN_3+5N_2$ to the predicted $K_2N_{16}$-P6/mcc compound upon hydrostatic compression to 80 GPa.

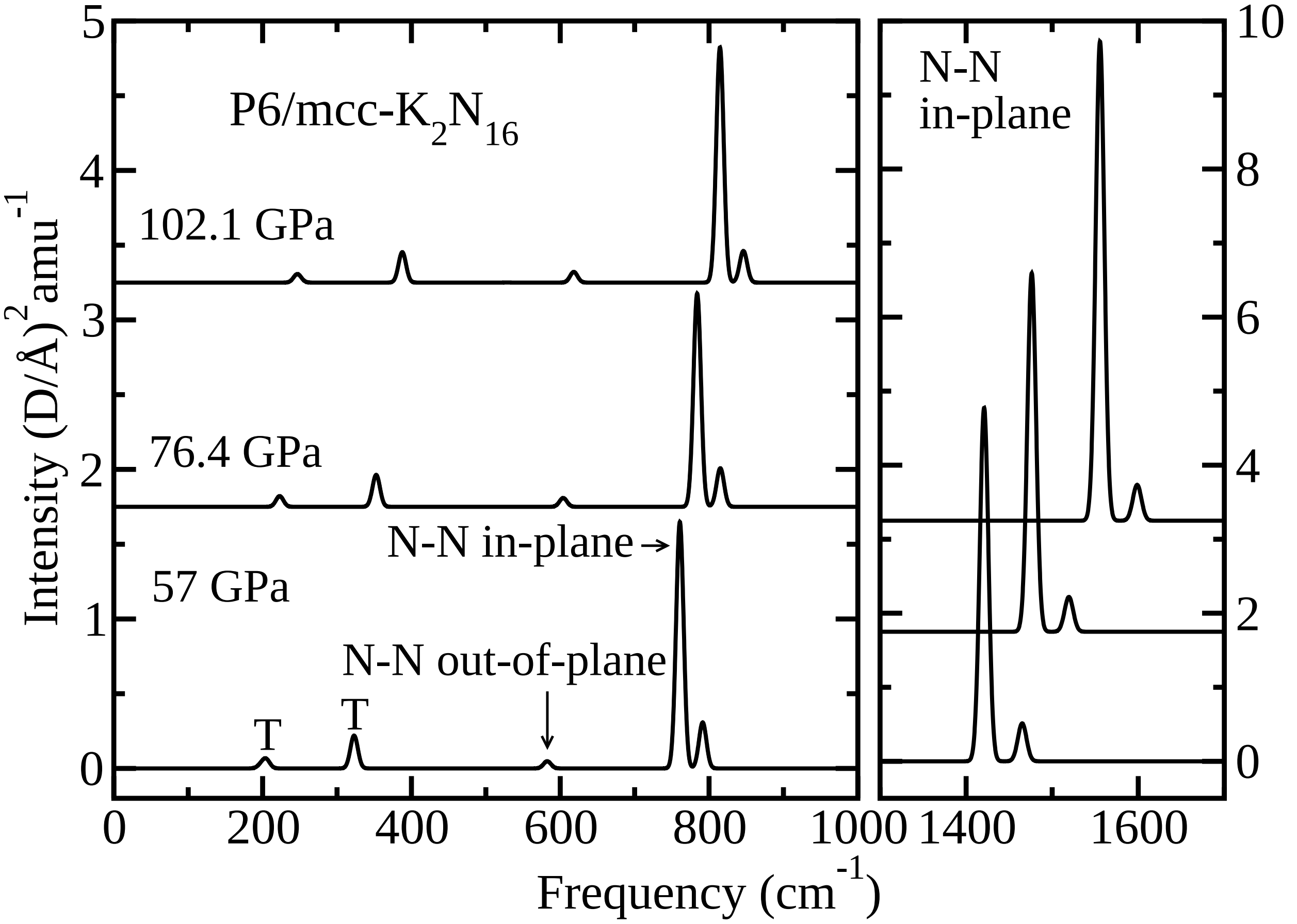

Figure 7: Calculated IR-spectra of the metallic layered nitrogen containing crystal $K_2N_{16}$-P6/mcc as a function of pressure. The spectra are split into two panels from 0-1,000 $cm^{-1}$ and 1,300-1,700 $cm^{-1}$, the second panel uses a larger scale to show the intensity at each pressure.

## Supporting Information Available

Relative enthalpy plot of structures with 1:1 ratio of K:N.

Cif files of nitrogen-rich structures. This material is available free of charge via the Internet at `http://pubs.acs.org/`.

## Acknowledgement

The research is supported by the Defense Threat Reduction Agency, (grant No. HDTRA1-12- 1-0023) and Army Research Laboratory through Cooperative Agreement W911NF-16-2-0022. Simulations were performed using the NSF XSEDE supercomputers (grant No. TG-MCA08X040), DOE BNL CFN computational user facility, and USF Research Computing Cluster supported by NSF (grant No. CHE-1531590).

**Corresponding Authors**

email: oleynik@usf.edu

**Notes**

The authors declare no competing financial interests.

# References

(1) Eremets, M. I.; Gavriliuk, A. G.; Trojan, I. A.; Dzivenko, D. A.; Boehler, R. Single-bonded cubic form of nitrogen. *Nat. Mat.* **2004**, *3*, 558–63.

(2) Tomasino, D.; Kim, M.; Smith, J.; Yoo, C. S. Pressure-induced Symmetry Lowering Transition in Dense Nitrogen to Layered Polymeric Nitrogen ( LP-N ) with Colossal Raman Intensity. *Phys. Rev. Lett.* **2014**, *113*, 205502.

(3) Christe, K. O. Recent Advances in the Chemistry of N5+, N5- and High-Oxygen Compounds. *Propellants, Explos., Pyrotech.* **2007**, *32*, 194–204.

(4) Eremets, M. I.; Gavriliuk, A. G.; Serebryanaya, N. R.; Trojan, I. A.; Dzivenko, D. A.; Boehler, R.; Mao, H. K.; Hemley, R. J. Structural transformation of molecular nitrogen to a single-bonded atomic state at high pressures. *J. Chem. Phys.* **2004**, *121*, 11296–300.

(5) Eremets, M. I.; Gavriliuk, A. G.; Trojan, I. A. Single-crystalline polymeric nitrogen. *Appl. Phys. Lett.* **2007**, *90*, 171904.

(6) Eremets, M. I.; Popov, M. Y.; Trojan, I. A.; Denisov, V. N.; Boehler, R.; Hemley, R. J. Polymerization of nitrogen in sodium azide. *J. Chem. Phys.* **2004**, *120*, 10618–23.

(7) Li, J.; Wang, X.; Xu, N.; Li, D.; Wang, D.; Chen, L. Pressure-induced polymerization of nitrogen in potassium azides. *EPL (Europhysics Letters)* **2013**, *104*, 16005.

(8) Peng, F.; Yao, Y.; Liu, H.; Ma, Y. Crystalline LiN5 Predicted from First-Principles as a Possible High-Energy Material. *J Phys Chem Lett* **2015**, *6*, 2363–6.

(9) Peng, F.; Han, Y.; Liu, H.; Yao, Y. Exotic stable cesium polynitrides at high pressure. *Sci Rep* **2015**, *5*, 16902.

(10) Shen, Y.; Oganov, A. R.; Qian, G.; Zhang, J.; Dong, H.; Zhu, Q.; Zhou, Z. Novel lithium-nitrogen compounds at ambient and high pressures. *Sci. Rep.* **2015**, *5*, 14204.

(11) Steele, B. A.; Oleynik, I. I. Sodium pentazolate: A nitrogen rich high energy density material. *Chem. Phys. Lett.* **2016**, *643*, 21–26.

(12) Steele, B. A.; Stavrou, E.; Crowhurst, J. C.; Zaug, J. M.; Prakapenka, V. B.; Oleynik, I. I. High-Pressure Synthesis of a Pentazolate Salt. *Chem. Mater.* **2017**, *29*, 735–741.

(13) Steele, B. A.; Oleynik, I. I. Pentazole and Ammonium Pentazolate: Crystalline Hydro-Nitrogens at High Pressure. *The Journal of Physical Chemistry A* **2017**, *121*, 1808–1813.

(14) Glass, C. W.; Oganov, A. R.; Hansen, N. USPEX-evolutionary crystal structure prediction. *Comp. Phys. Comm.* **2006**, *175*, 713–720.

(15) Lyakhov, A. O.; Oganov, A. R.; Valle, M. How to predict very large and complex crystal structures. *Comp. Phys. Comm.* **2010**, *181*, 1623–1632.

(16) Oganov, A. R.; Glass, C. W. Crystal structure prediction using ab initio evolutionary techniques: principles and applications. *J. Chem. Phys.* **2006**, *124*, 244704.

(17) Kresse, G.; Furthmiiller, J. Efficiency of ab-initio total energy calculations for metals and semiconductors using a plane-wave basis set. *Comp. Mat. Sci.* **1996**, *6*, 15–50.

(18) Pickard, C. J.; Needs, R. J. Ab initio random structure searching. *J. Phys. Cond. Matt.* **2011**, *23*, 053201.

(19) Ji, C.; Zhang, F.; Hou, D.; Zhu, H.; Wu, J.; Chyu, M.-C.; Levitas, V. I.; Ma, Y. High pressure X-ray diffraction study of potassium azide. *J. Phys. Chem. Solids* **2011**, *72*, 736–739.

(20) Ji, C.; Zheng, R.; Hou, D.; Zhu, H.; Wu, J.; Chyu, M. C.; Ma, Y. Pressure-induced phase transition in potassium azide up to 55 GPa. *J. Appl. Phys.* **2012**, *111*, 112613.

(21) Perdew, J.; Burke, K.; Ernzerhof, M. Generalized Gradient Approximation Made Simple. *Phys. Rev. Lett.* **1996**, *77*, 3865–3868.

(22) Grimme, S. Semiempirical GGA-Type Density Functional Constructed with a Long-Range Dispersion Correction. *J. Comp. Chem.* **2006**, *27*, 1787–1799.

(23) Kresse, G.; Joubert, D. From ultrasoft pseudopotentials to the projector augmented-wave method. *Phys. Rev. B* **1999**, *59*, 1758–1775.

(24) Lazzeri, M.; Mauri, F. First-Principles Calculation of Vibrational Raman Spectra in Large Systems: Signature of Small Rings in Crystalline SiO2. *Phys. Rev. Lett.* **2003**, *90*, 036401.

(25) Porezag, D.; Pederson, M. R. Infrared intensities and Raman-scattering activities within density-functional theory. *Phys. Rev. B* **1996**, *54*, 7830–7836.

(26) Delley, B. From molecules to solids with the DMol3 approach. *J. Chem. Phys.* **2000**, *113*, 7756.

(27) Zhang, J.; Zeng, Z.; Lin, H.-Q.; Li, Y.-L. Pressure-induced planar N6 rings in potassium azide. *Sci Rep* **2014**, *4*, 4358.

(28) Zhang, M.; Yin, K.; Zhang, X.; Wang, H.; Li, Q.; Wu, Z. Structural and electronic properties of sodium azide at high pressure: A first principles study. *Solid State Comm.* **2013**, *161*, 13–18.

(29) McMahan, A. K. Alkali-metal structures above the s-d transition. *Phys. Rev. B* **1984**, *29*.

(30) Mailhiot, C.; Yang, L.; McMahan, A. Polymeric nitrogen. *Phys. Rev. B* **1992**, *46*, 14419–14435.